\documentclass[aps,prd,reprint,superscriptaddress,twocolumn,preprintnumbers,nofootinbib]{revtex4-2}
\usepackage{hyphenat}
\usepackage{comment}
\usepackage[utf8]{inputenc} 
\usepackage{graphicx,color,overpic,mathtools}
\usepackage{amsthm,amsmath,amssymb,hyperref,mathrsfs}
\usepackage{mathrsfs,bm}
\usepackage{natbib,textcase}
\usepackage{flexisym}
\usepackage{braket,bbm,setspace}
\PassOptionsToPackage{normalem}{ulem}
\usepackage{ulem} 
\usepackage{physics}
\usepackage{float}
\usepackage{caption}
\usepackage[makeroom]{cancel}
\usepackage[english]{babel}
\graphicspath{ {images/} }
\usepackage{breqn}

\addto\captionsspanish{}

\usepackage[usenames,dvipsnames]{xcolor}

\def\be#1\ee{\begin{align}#1\end{align}}

\renewcommand{\ge}{\geqslant}
\renewcommand{\geq}{\geqslant}

\newcommand{\ee}[1]{\end{equation}}
\makeatletter
\let\cat@comma@active\@empty
\makeatother

\begin{document}

\title{
From de Sitter to Anti-de Sitter singularity regularization:\\ theory and phenomenology
}

\author{Julio Arrechea}
\email{julio.arrechea@sissa.it}
\author{Stefano Liberati}
\email{liberati@sissa.it}
\author{Hooman Neshat}
\email{hneshat@sissa.it}
\affiliation{SISSA, via Bonomea 265, 34136 Trieste, Italy.}
\affiliation{INFN (Sez. Trieste), via Valerio 2, 34127 Trieste, Italy.}
\affiliation{IFPU, via Beirut 2, 34014 Trieste, Italy.}
\author{Vania Vellucci}
\email{vellucci@qtc.sdu.dk}
\affiliation{Quantum Theory Center ($\hbar$QTC) \& D-IAS, IMADA at Southern Denmark Univ.,\\ Campusvej 55, 5230 Odense M, Denmark.}

%----------------------------------------------------------
\begin{abstract}
{Recent investigations of vacuum polarization in extremely compact stars suggest that, in such regimes, the effective matter content of spacetime may acquire a vacuum-energy equation of state with negative energy density, mimicking a negative cosmological constant. Motivated by this observation, we introduce a general algorithm to modify well-known spherically symmetric regular black hole metrics by replacing their usual de Sitter cores (dSC) with Anti–de Sitter cores (AdSC). Like their dSC counterparts, these AdSC solutions may exhibit two, one, or no horizons depending on the value of a regularization parameter $\ell$. We present explicit examples of \mbox{AdSC–Bardeen} and \mbox{AdSC–Dymnikova} metrics, analyze their main properties, and investigate some of their phenomenological signatures using test fields. In particular, we compare their fundamental quasinormal modes and echo signals with those of the dSC cases, highlighting potential avenues for distinguishing them observationally.}
\end{abstract}

\maketitle

\section{Introduction}
{The search for regular alternatives to General Relativity (GR) black holes has opened a novel field of investigation regarding the structure and the phenomenological properties of non-GR black holes and horizonless ultra compact objects. These objects, whose existence can be postulated from physical principles of varied origins~\cite{MazurMottola2004,Mathur2005,Visseretal2009,HoldomRen2017,CardosoPani2019,Bambi:2025wjx}, provide alternatives (devoid of some of the pathologies associated to GR black holes) to the dark and compact objects that, according to recent gravitational-wave~\cite{LIGOScientific2016, LIGOScientific2017} and long-baseline-interferometry observations~\cite{EventHorizonTelescope2019,EventHorizonTelescope2022}, are ubiquitous in our universe.

In some cases, these alternatives to GR singular black holes can be seen as solutions of the standard Einstein equations in the presence of an effective stress-energy tensor associated with some exotic (specifically strong energy condition violating) matter, most often inspired from non-linear electrodynamics~\cite{Ayon-Beato:2000mjt,Bronnikov:2022ofk}. Alternatively, they can be derived as vacuum, albeit non-Ricci flat, solutions of some modified gravity theories (see e.g.~\cite{Bueno:2024dgm, Bueno:2024eig, Boyanov:2025otp}). Or even solutions coupled to Maxwell and nonlinear electrodynamics, in theories incorporating infinite towers of higher-order curvature corrections \cite{Hennigar:2025ftm}.

Depending on which features from the standard black hole paradigm these models maintain, we can broadly divide current proposals into those which exhibit horizons, such as regular black holes~\cite{Bardeen1968,Dymnikova:1992ux,Hayward2006,Ansoldi:2008jw}, and those which are horizonless (see e.g.~\cite{Mathur2005,Arrecheaetal2022}) but still so compact to reproduce most of the black hole external features (e.g.~their shadow), hence the name of black hole mimickers~\cite{Carballo-Rubio:2025fnc}.\footnote{Note that by this name, we will be referring to horizonless objects that would replace GR black holes of any mass, and not just coexist with them as an additional family of compact objects. Examples of the latter would be boson and Proca stars~\cite{Colpietal1986,Britoetal2015}, which we leave aside our discussion here.}
In this article, we present new families of regular black holes and black hole mimickers which exhibit inner regions with negative energy densities, a physically viable possibility that has been, for the most part, overlooked in previous literature. Let us first give a brief overview on these models.

\subsection{Regular black holes vs black hole mimickers}
Regular black holes (RBHs) fall broadly in two families depending on the way the GR singularity is regularized~\cite{Carballo-Rubio:2019nel,Carballo-Rubio:2019fnb}. In one case, the interior displays a minimum radius beyond which there is an anti-trapped region --- these are often called black bounces or hidden wormhole black holes (see e.g.~\cite{SimpsonVisser2018,Mazzaetal2021}) --- in the other case the interior displays a locally untrapped region, i.e.~a region where the outgoing null rays can move outward. Of course, for such a region to exist beyond a proper outer trapping horizon, an inner horizon must exist. 
%In stationary metrics such inner horizons are also a Cauchy horizon. 

The two classes have quite different properties and features. As such, they deserve separate treatments. In what follows, we shall deal only with the second class of RBHs, limiting ourselves to spherically symmetric metric, and leave the first to future investigations. 

RBHs with untrapped cores indeed have been the most studied in the literature and those introduced earlier (see \cite{Bardeen1968}). This is partly because they can be recovered as a minor modifications of GR stationary black hole metrics: by just replacing the ADM mass $M$ with $m(r)$, the Misner--Sharp--Hernandez mass~\cite{MisnerSharp1964,HernandezMisner1966} and requiring that the latter goes to zero sufficiently fast for $r\to 0$, so as to ensure regularity of all the curvature scalars.

In spite of their relative conservative nature, such RBHs, when stationary, are nonetheless well known to be subject to instabilities of both classical (mass inflation)~\cite{Poisson:1989zz,Carballo-Rubio:2024dca}, and semiclassical~\cite{BalbinotPoisson1993,Hollands:2019whz,Barceloetal2022,McMaken:2023uue} nature. Both of these instabilities are traceable to the fact that the inner horizons bounding such untrapped cores are also Cauchy horizons, and indeed are present also in GR black holes sporting the same structures (e.g. Reissner--Nordstr\"om or Kerr black holes). This implies that such spacetimes cannot be stationary, and in general they might be rapidly evolving.\footnote{Noticeably, for slowly evolving inner horizons (which are not Cauchy horizons) the above mentioned classical instability was still found to exist~\cite{Carballo-Rubio:2024dca}.} {Recent analyses even display a disappearance of the trapped region when semiclassical backreaction effects are included~\cite{Barenboim:2025ckx,Boyanov:2025otp}.}

Remarkably, one-parameter families of metrics describing RBHs can also describe horizonless objects~\cite{Carballo-Rubioetal2022b}, as long as the value of their regularization parameters (which naively would estimate the scale of quantum-gravitational corrections) is comparable to the mass of the system instead of Planckian. The above mentioned instabilities might be the mechanism by which the regularization parameter might acquire a time dependence and grow to such large values, so evolving the RBH into an horizonless black hole mimicker.\footnote{Note, however, such horizonless alternatives, are not devoid of problems either, being subject to potential instabilities associated with the presence of stable light rings~\cite{Cardoso:2014sna,Cunha:2022gde,DiFilippo:2024ddg}, ergoregions~\cite{Maggio:2017ivp}, and accretion~\cite{Addazi:2019bjz,Vellucci:2022hpl}.  It is still unclear how fast and which kind of evolution such additional instabilities could drive.} 
%In summary, the regularization of singularities seems to unlock a very rich physics and naturally uncover to a quite dynamical nature of the resulting objects. Addressing the formation and stability of the latters would be crucial in advancing this field of investigation, but it will require understanding dynamics beyond GR (see e.g. for recent attempts~\cite{Barenboim:2025ckx,Boyanov:2025otp,Boyanov:2025pes}) or within GR but coupling gravity to non-conventional matter sectors (see e.g.~\cite{Electrodynamics}).\\
Nevertheless, in this work we shall put aside issues concerning the dynamical evolution of RBHs and black hole mimickers, to rather focus on stationary geometries associated with this class of objects when an Anti-de Sitter core is present. 

\subsection{Anti-de Sitter cores}

We already mentioned that a simple procedure, to generate a RBH geometry from a GR one (especially for static, spherically symmetric geometries), consists in replacing its ADM mass with a Misner--Sharp--Hernandez one which goes sufficiently fast to zero. More precisely, for the standard Schwarzschild metric element one gets, $g_{tt}=(g_{rr})^{-1}=(1-2M/r)\to (1-2m(r)/r)$, and all one enforces is $m(r)\sim r^3$ as $r\to 0$ to ensure a regular behavior of all the relevant curvature scalars at the origin (see e.g.~\cite{Torres:2016pgk} for a review). Now, it is easy to realize that this implies $g_{tt}=(g_{rr})^{-1}\to (1-A\, r^2)$ for $r\to 0$, where $A$ is some non-zero constant. So, depending on the sign of $A$, a de Sitter (dS) or Anti-de Sitter (AdS) metric is generically recovered at the core of these black hole spacetimes.

In the extant literature, most of the spherically symmetric metrics studied are characterized by de Sitter cores. The reason for this bias is rooted in the history of the RBH paradigm, which stemmed from early ideas about cosmology and the regularization of the big bang singularity (see e.g.~\cite{GlinerDymnikova1974,Starobinsky:1980te, Brandenberger:1993ef}). Such ideas were later exported to black holes (see~\cite{Bardeen1968, Frolov:1988vj, Dymnikova:1992ux, Poisson:1988wc, Elizalde:2000xn, Hayward2006, Ansoldi:2008jw} and references therein) by adopting the same underlying logic: if matter suffers gravitational collapse, densities reaching a Planckian threshold may trigger a phase transition into a new, vacuum-like phase~\cite{Gliner:1966cgu,Markov1982}. For what concerns the equation of state of such phase, singularity avoidance requires the violation of some energy condition, generally the strong energy condition (SEC), which for a perfect fluid takes the form: $\forall i$, $\rho + p_i \ge 0$ and $\rho + \sum_i p_i \ge 0$~\cite{Curiel2017}. Since the quantity that triggers this transition is the density $\rho$ --- which needs to surpass a certain \textit{positive} threshold scale --- the violation of the SEC requires (some) of the pressures $p_{i}$ to become \textit{negative}. 

A further reinforcement to the above physical picture came from the gravastar framework~\cite{MazurMottola2004}. In~\cite{MazurMottola2015} it was shown how the gravastar model arises naturally, within GR, from increasing the compactness of constant-density fluid spheres beyond their maximum compactness limit (the Buchdahl limit~\cite{Buchdahl1959}) all the way to the black hole limit. This derivation assumed the validity of GR as an effective theory across a surface of infinite curvature, beyond which the interior of the star resembles de Sitter spacetime. Below we present the alternative scenario that emerges when vacuum polarization effects are taken into account.

In considering quantum field theory corrections to stellar equilibrium it was recently shown that the gravitational contributions of the quantum vacuum need to be taken into account as the Buchdahl limit is approached~\cite{Arrecheaetal2022,Arrechea:2023oax,Reyes:2023fde,Reyes:2023ags}.  Indeed, it was observed that as the star is compressed and made to approach the Buchdahl limit, its internal (classical) pressures become arbitrarily large and positive, acting as the seed for subsequent large vacuum polarization effects that take the form of an additional anisotropic stress-energy tensor, whose backreaction must be taken into account.

Eventually, a regime is reached in which the polarization of the vacuum becomes dominant in the innermost regions of the star, and the effective energy density (the sum of classical and semiclassical contributions) becomes negative. In this situation, what triggers the transition of the effective matter content towards the vacuum equation of state is the crossing of a certain \textit{positive} pressure threshold, and what regularizes the singularity is the emergence of an effective density that in this limit becomes \textit{negative}. Figure~\ref{Fig:Gravastars} shows a pictorial comparison between the gravastar and the semiclassical star scenarios.
\begin{figure}
    \centering
    \includegraphics[width=\linewidth]{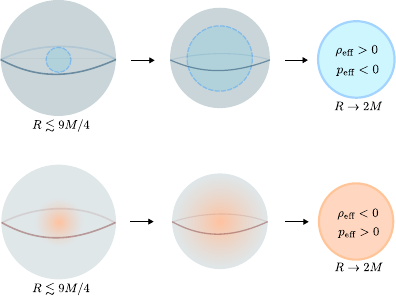}
    \caption{Top panel: Sequence of constant-density stars approaching a gravastar in the black hole limit. As the star is compressed, a spherical shell of infinite curvature moves towards the surface, inside of which lies a region of negative effective pressure (in blue). Bottom panel: Sequence of constant-density stars approaching the black hole limit together with their vacuum polarization effects. The star is regular and displays a core of negative effective energy that expands to fill the whole interior in the black hole limit, where they can be approximated by the AdS counterparts of gravastars~\cite{Arrechea:2024vxp}.}
    \label{Fig:Gravastars}
\end{figure}

These semiclassical stellar solutions display a broad core with negative mass surrounded by a narrow crust of positive mass which resembles, in a most idealised description, a gravastar with an Anti-de Sitter interior. They can be as compact as black holes and act as very efficient black hole mimickers due to their large internal light-crossing times~\cite{Arrechea:2024nlp}.Interestingly, similar models have been considered inspired by string theory~\cite{Danielsson:2017riq,Danielsson:2021ykm,Giri:2024cks}, as AdS counterparts to gravastars themselves~\cite{Adler:2022fqu, Adler:2023tpi}, or as particular bilayered constant-density stars~\cite{Arrechea:2024vxp}. 

In both scenarios (gravitational collapse and stellar equilibrium) the effective matter content evolves towards a vacuum equation of state, but the signs of the energy and pressures are swapped depending on the way that the system is allowed to reach the degree of compression required to trigger such transition. In the end, it is very likely that vacuum polarization effects can give rise to both AdS and dS equations of state, and it may simply depend on the properties of the system—--namely, whether density or pressure dominates—--which scenario is realized. RBHs allow to play with both possibilities as they can be considered to describe regular collapse processes or horizonless equilibrium configurations, so we shall use them here as a tool to explore the consequences of having AdS interiors.

The plan of this work is the following. 
%In section~\ref{Sec:physmot} we provide a more articulated physical motivation for AdS core black holes and black hole mimickers.  
Section~\ref{Sec:RBHs} introduces a prescription to generate ``AdS versions" of known RBH metrics. We explicitly discuss its application to the Bardeen and Dymnikova metrics, and study the shape of the effective stress-energy tensor sourcing them. Section~\ref{Sec:Pheno} compares the behavior of the fundamental quasinormal mode (QNM) of the Bardeen and Dymnikova standard metrics with their Anti-de Sitter counterparts. We also study the time evolution of a test spin-2 field for the horizonless counterparts of these AdSC metrics, in particular the ``echoes" associated with the absence of long internal light-crossing times.
Finally, in Section~\ref{Sec:Conclusions} we shall draw our conclusions and discuss possible avenues for further advancement of our investigation.

\section{Regular black holes with Anti-de Sitter cores}
\label{Sec:RBHs}

The Einstein equations 
 $G^{\mu}_{~\nu}=8\pi T^{\mu}_{~\nu}$
relate the curvature of spacetime, given by the Einstein tensor $G_{\mu\nu}$, with its stress-energy content, given by $T_{\mu\nu}$. While GR itself is a classical theory, in semiclassical gravity $T_{\mu\nu}$ can become an effective stress-energy tensor, describing not just classical matter but quantum effects associated with the dynamical degrees of freedom of the theory (including gravitons). This typically translates into a $T_{\mu\nu}$ that can violate some if not all of the energy conditions and that can be used as a collective description of physics beyond GR. 

Equivalently, the Einstein equations always allow us to use a chosen metric as an input and derive (by computing the related Einstein tensor $G_{\mu\nu}$) the corresponding effective stress-energy content necessary for generating such a spacetime, which again can be seen as originating in some unknown extension of GR. As anticipated in the Introduction, several efforts are under way to derive RBH metrics (and/or their horizonless counterparts) from alternative theories of gravity. In what follows, we shall adopt an agnostic viewpoint on the physical origin of the proposed models. 

We restrict ourselves to eternal and spherically symmetric spacetimes described by the line element
\begin{equation}\label{Eq:SphSymmetric}
    ds^{2}=-f(r)dt^{2}+h(r)dr^{2}+r^{2}d\Omega^{2}\, ,
\end{equation}
where $d\Omega^{2}$ is the line element of the unit $2-$sphere. In these coordinates, the stress-energy tensor measured by static observers can be written in diagonal form as
\begin{equation}    T^{\mu}_{~\nu}=\text{diag}\left(-\rho,p_{r},p_{\theta},p_{\theta}\right)\, ,
\end{equation}
where $\rho(r)$ denotes the energy density and $p_{r}(r)$ and $p_{\theta}(r)$ are the pressures in the radial and angular directions, respectively. 
The components of the relevant Einstein equations are 
\begin{align}\label{Eq:Einsteinsphsym}
    &
    -\frac{1}{r^{2}}+\frac{1}{r^{2}h}-\frac{h'}{rh^{2}}=
    -8\pi \rho\,,\nonumber\\
    &
    -\frac{1}{r^{2}}+\frac{1}{r^{2}h}+\frac{f'}{rfh}=
    8\pi p_{r}\,,\nonumber\\
    &
    \frac{f'}{2rfh}-\frac{(f')^{2}}{4f^{2}h}-\frac{h'}{2rh^{2}}-\frac{f'h'}{4fh^{2}}+\frac{f''}{2fh}=
    8\pi p_{\theta}\
\end{align}

Typical RBH models have simple metrics of the form \mbox{$f(r)=h(r)^{-1}=1-2m(r)/r$}. It is easy to see (using the first two equations in \eqref{Eq:Einsteinsphsym}) that such metric ansatz implies
%Throughout the Einstein equations, this constraint \note{Which constraints?!?!?} translates into 
an effective SET satisfying $\rho+p_{r}=0$ globally. As mentioned in the Introduction, this equation of state can be realized with either positive or negative $\rho$ depending whether one considers collapse or equilibrium situations. For what concerns RBH spacetimes, they are allowed to exhibit regions with negative $\rho$, as long as the Misner--Sharp--Hernandez mass is positive at infinity.
%Any two equations in~\eqref{Eq:Einsteinsphsym} together with the conservation relation
%\begin{equation}
%\nabla_{\mu}T^{\mu}_{~r}=p'+\frac{f'}{2f}\left(\rho+p_{r}\right)+\frac{2}{r}\left(p_{r}-p_{\theta}\right),
%\end{equation}
%form a closed system of equations.
%Stress-energy tensors of such simple forms can be source by non-linear electrodynamic models~\cite{}, which provide a physical theory onto which study their dynamical aspects such as their stability~\cite{}.
In what follows, we shall identify the core of a RBH as de Sitter if the effective SET obeys
\begin{equation}
    \rho=-p_{r}=-p_{\theta}=\frac{\Lambda_{0}}{8\pi}+\order{r^{2}}, \quad \Lambda_{0}>0\, ,
\end{equation}
and as Anti-de Sitter if $\Lambda_{0}<0$. 

It is possible to obtain Anti-de Sitter-core versions of de Sitter-core RBHs by decomposing the mass function into
\begin{equation}\label{Eq:fAdSC}
    f=1-\frac{2m(r)}{r}=1-\frac{2m_{\rm dS}(r)\chi(r)}{r}\, .
\end{equation}
Here, $m_{\rm dS}$ is the Misner--Sharp--Hernandez mass of the de Sitter-Core solution and $\chi(r)$ is any function that interpolates between the asymptotic values
\begin{equation}
\lim_{r\to0}\chi=-1+\order{r^\alpha},\quad \lim_{r\to\infty}\chi=1+\order{r^{-\beta}}\, ,
\end{equation}
with $\alpha,~\beta$ some positive constants.
Among all the possible choices of $\chi$ function, we select the family 
\begin{equation}\label{Eq:ChiRBH}
    \chi(r,n,\epsilon)=\frac{r^{n}-\epsilon \ell^{n}}{r^{n}+ \epsilon\ell^{n}},\quad n\geq 1\, ,
\end{equation}
where $\epsilon$ is some positive constant. In most RBHs it will be left equal to $1$, but we can take $0<\epsilon\ll 1$ to obtain AdSC versions of RBHs that are arbitrarily close to their dSC counterparts, except for an arbitrarily small central region.
Since we are interested in magnifying the differences between these two types of cores, we stick to the choice~\eqref{Eq:ChiRBH} with $\epsilon=1$ throughout the rest of the paper unless stated otherwise.
Similarly, $\ell$ denotes the regularization length scale that typically appears in regular black hole solutions. While it might naively be associated with the Planck scale, $\ell$ need not coincide with it. For instance, the quantum–gravitational effects responsible for regularizing the singularity may become relevant at the Planck density rather than at a Planck radius, leading to a different characteristic length. In addition, $\ell$ may transiently evolve to larger values in dynamical scenarios due to instabilities known to arise near inner horizons. For these reasons, in this work we treat $\ell$ as a free parameter, allowing us to explore a broader class of regular black hole and black hole mimicker geometries.
%\footnote{\jar{Some of the AdSC generalizations that we propose, as is the case for the AdSC-Dymnikova metric proposed below~\eqref{Eq:fDymnikovaAdS}, are not generated through a prescription such as Eq.~\eqref{Eq:fAdSC}, since the function $\xi$ would spoil the exponential decay approach of the Dymnikova mass function towards the ADM mass.}}
Let us now discuss in detail the AdSC version of some of the best-known RBHs spacetimes. 
Throughout this work, we employ units with $G=c=1$, and we set the ADM mass to $M=1$ unless explicitly stated otherwise.

\subsection{Bardeen metric}
\subsubsection{Metric properties}
The Bardeen mass function~\cite{Bardeen1968} is given by
\begin{equation}
    m_{\scriptscriptstyle \text{B--dS}}=M\left(\frac{r}{\sqrt{r^2+\ell ^2}}\right)^{3}\, .
\end{equation}
An expansion around $r=0$ of this mass function shows
\begin{equation}
    \lim_{r\to0} m_{\scriptscriptstyle \text{B--dS}}=Mr^{3}/\ell^{3},\quad \lim_{r\to0}f_{\scriptscriptstyle \text{B--dS}}=1-2M r^{2}/\ell ^{3}\,.
\end{equation}
Identifying $\Lambda_{0}=6M/\ell^{3}$ we see the above metric is of the static de Sitter form near $r=0$.

The Anti-de Sitter version of this metric can be generated by the previously described algorithm, for example taking the simplest case $\chi(r,2,1)$ (let us stress that this is just one of many possibilities), so that
\begin{equation}
    m_{\scriptscriptstyle \text{B--AdS}}=M\left(\frac{r}{\sqrt{r^2+\ell ^2}}\right)^{3}\left(\frac{r^{2}-\ell^{2}}{r^{2}+\ell^{2}}\right)\,.
\end{equation}
%where we have taken $\{n,\epsilon\}=\{2,1\}$ in~\eqref{Eq:ChiRBH}.
In this way, at the core we have
\begin{equation}\label{Eq:Bardeenr0}
    \lim_{r\to 0}m_{\scriptscriptstyle \text{B--AdS}}=-Mr^{3}/\ell^{3},\quad \lim_{r\to0}f_{\scriptscriptstyle \text{B--AdS}}=1+2M r^{2}/\ell ^{3}\,.
\end{equation}

Similarly to their dSC version, these AdSC solutions exhibit two, one, or zero horizons depending on the value of the regularization parameter $\ell$. Figure~\ref{Fig:Bardeenf} shows a plot of the metric function $f$ for various values of $\ell$. In the $\ell\to0$ and $\ell\to\infty$ limits this metric approaches the Schwarzschild and Minkowski spacetimes, respectively.
\begin{figure}
    \centering
    \includegraphics[width=\linewidth]{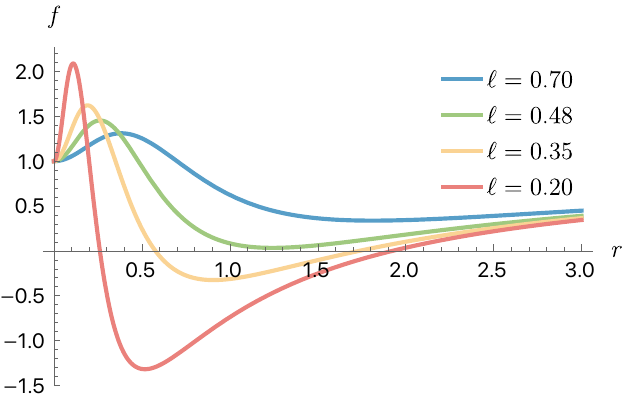}
    \caption{Plot of the metric function $f_{\scriptscriptstyle \text{B--AdS}}$ for different values of the regularization parameter $\ell$. Depending on its value, the metric can exhibit two, one, or no horizons. The Anti-de Sitter core causes $f$ to take values greater than $1$ at small radii.}
    \label{Fig:Bardeenf}
\end{figure}
Notice that there will be a critical value of $\ell$ for which the geometry exhibits a single, extremal horizon, and that this value of $\ell$ differs from the corresponding one in the dSC-Bardeen metric. Similarly, these metrics exhibit an interior with negative mass surrounded by an exterior of positive mass. The way the mass approaches its asymptotic value also depends on which core is chosen, i.e.
\begin{equation}
    \lim_{r\to\infty}m_{\scriptscriptstyle \text{B--dS}}=M-\frac{3M\ell^2}{2r^2},\quad \lim_{r\to\infty}m_{\scriptscriptstyle \text{B--AdS}}=M+\frac{7M\ell^2}{2r^2}\,.
\end{equation}

Figure~\ref{Fig:BardeenAdSdS} shows a comparison between the metric functions of the dSC-Bardeen and AdSC-Bardeen spacetimes, where we see the largest departures between both spacetimes take place in the innermost regions. Finally, the Ricci scalar in the coordinates~\eqref{Eq:SphSymmetric}
\begin{align}\label{Eq:Rscalar}
    \mathcal{R}=
    &
    \frac{2}{r^2}\left(1-\frac{1}{h}\right)+\frac{2}{hr}\left(\frac{h'}{h}-\frac{f'}{f}+\frac{rf'h'}{4fh}\right)\nonumber\\
    &
    +\frac{1}{2h}\left[\left(\frac{f'}{f}\right)^{2}-\frac{2f''}{f}\right].
\end{align}
obeys the expansions
\begin{align}
\lim_{r\to0}\mathcal{R}_{\scriptscriptstyle \text{B--AdS}}=
&
-\frac{24M}{\ell^3}+\mathcal{O}\left(r^2\right),\nonumber\\ \lim_{r\to\infty}\mathcal{R}_{\scriptscriptstyle \text{B--AdS}}=
&
-\frac{14M \ell^2}{r^5}+\mathcal{O}\left(r^{-6}\right).
\end{align}
Notice the change in sign of the Ricci scalar with respect to the Bardeen metric, where $\mathcal{R}_{\scriptscriptstyle \text{B--dS}}(r=0)=24M/\ell^3$.
\begin{figure}
    \centering
    \includegraphics[width=\linewidth]{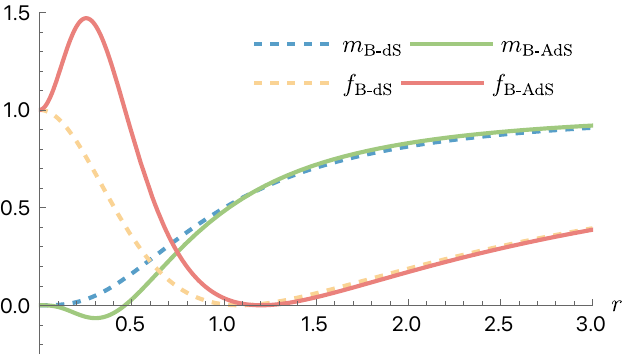}
    \caption{Comparison of the $f$ and $m$ functions for the dSC and AdSC-Bardeen metrics. The regularization parameters have been fixed to $\ell_{\scriptscriptstyle \text{B--dS}}\approx 0.77$ and $\ell_{\scriptscriptstyle \text{B--AdS}}\approx 0.46$ so that both solutions exhibit a single, extremal horizon, even though this horizon is at different radial positions for both models. While their geometries are similar at large $r$, discrepancies appear near the origin. In particular, the Misner--Sharp--Hernandez mass takes negative values in some region for the AdSC-Bardeen metric.}
    \label{Fig:BardeenAdSdS}
\end{figure}

\subsubsection{Stress-energy tensor}
The components of the stress-energy source giving rise to the AdSC-Bardeen metric can be obtained from the Einstein equations~\eqref{Eq:Einsteinsphsym}, namely,
\begin{align}
    \rho=-p_{r}=
    &
    \frac{M \ell^{2}\left(7r^{2}-3\ell^{2}\right)}{4\pi\left(r^{2}+\ell^{2}\right)^{7/2}}\nonumber\\
    p_{\theta}=
    &
    \frac{M\ell^{2}\left(21r^{4}-43r^{2}\ell^{2}+6\ell^{4}\right)}{8\pi\left(r^{2}+\ell^{2}\right)^{9/2}}.
\end{align}
Figure~\ref{Fig:SETBardeen} shows these components for a particular value of $\ell$ that corresponds to a horizonless object. The most significant feature in the SET of these solutions is the inversion in the change in sign of the energy density as a requisite to have an AdS core. This change of sign is similar to the behavior of the total energy density in semiclassical stellar solutions~\cite{Arrechea:2023oax}. The angular pressure displays a richer structure as it is positive in the exterior, negative in an intermediate region, and positive again near the origin. 
\begin{figure}
    \centering
    \includegraphics[width=\linewidth]{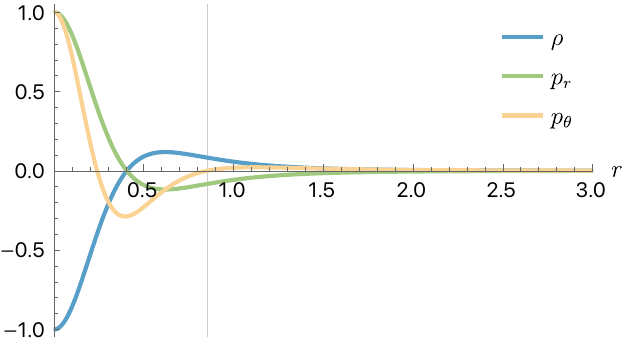}
    \caption{Stress-energy tensor components of the AdSC-Bardeen solution for $\ell=0.62$, which corresponds to a solution of the horizonless kind. The negative energy interior must be surrounded by an exterior region of positive energy density to ensure positivity of the mass at infinity. For this particular solution, the strong energy condition is violated inside a sphere of radius $r\approx 0.854$, indicated by the thin vertical line.}
    \label{Fig:SETBardeen}
\end{figure}

The radius where $\rho$ changes sign is
\begin{equation}
    r_{\rho}=\sqrt{\frac{3}{7}}\ell\, ,
\end{equation}
hence the weak and dominant energy conditions will be always violated within a sphere of radius $r_{\rho}$ where $\rho$ is negative. Since $\rho+p_{r}=0$ by construction, the strong energy condition is everywhere saturated if
\begin{equation}
    \rho+p_{\theta}\geq0\,,\quad p_{\theta}\geq0\,. 
\end{equation}
The first inequality is violated for $0<r< \ell$, while the second one holds everywhere but in the region \mbox{$b_{-}\ell<r<b_{+}\ell$}, with $b_{\pm}=\sqrt{\frac{43\pm\sqrt{1345}}{42}}$. Thus the strong energy condition is always violated inside a sphere of radius $b_{+}\ell$. For RBHs, this provides the necessary condition for violating the Hawking-Penrose singularity theorem~\cite{Hawking1970}. Convergence conditions associated to circumventing singularity theorems have been analyzed in detail for the AdSC-Bardeen spacetime in~\cite{Borissova:2025hmj} (see also~\cite{Borissova:2025msp} for dSC metrics), and recent works have also considered the vacuum polarization effects of the horizonless Bardeen spacetime under some approximations~\cite{Numajiri:2024qgh,Boasso:2025ofd}. 

\subsection{Dymnikova metric}
\subsubsection{Metric properties}
Among the variety of RBH models commonly considered, the Dymnikova metric~\cite{Dymnikova:1992ux} is of particular phenomenological interest, as it is derived from an effective SET whose components reproduce a de Sitter behavior near $r\to0$ and decay exponentially at large radii. Consequently, in the regime $\ell\ll M$, the outer horizon and the light ring lie at radial positions very close to those of the Schwarzschild black hole.
Explicitly, the metric function reads
\begin{equation}\label{Eq:fDymnikovadS}
    f_{\scriptscriptstyle \text{D--dS}}=1-\frac{2M}{r}\left[1-e^{-{r^{3}}/{\ell^{3}}}\right]\, .
\end{equation}

However, a problem arises here: applying our prescription in Eq.~\eqref{Eq:fAdSC} would make the function $\chi(r)$ spoil the exponential approach of the Dymnikova mass function to the ADM mass.

To preserve this phenomenologically appealing behavior, we therefore adopt a slightly modified version of our prescription; specifically, we apply the function $\chi(r,n,\epsilon)$ only in the exponential of the Dymnikova metric element. More precisely, we take~\eqref{Eq:ChiRBH} with $\chi(r,n,\epsilon)=\chi(r,2,1)$ to obtain an AdSC–Dymnikova metric with metric element of the form
\begin{equation}\label{Eq:fDymnikovaAdS}
    f_{\scriptscriptstyle \text{D--AdS}}=1-\frac{2M}{r}\left[1-\exp{-\frac{r^{3}}{\ell^{3}}\left(\frac{r^{2}-\ell^{2}}{r^{2}+\ell^{2}}\right)}\right]\, .
\end{equation}

It is easy to see, that for $r\to0$ this function obeys identical expansions to~\eqref{Eq:Bardeenr0} while the Ricci scalar behaves as
\begin{align}
\lim_{r\to0}\mathcal{R}_{\scriptscriptstyle \text{D--AdS}}=
&
-\frac{24M}{\ell^3}+\mathcal{O}\left(r^2\right)\,,\\ 
\lim_{r\to\infty}\mathcal{R}_{\scriptscriptstyle \text{D--AdS}}=
&
e^{-r^3/\ell^3+\mathcal{O}(r^{-2})}\left(-\frac{18  M r^3}{\ell^6}+\mathcal{O}(r)\right)\,.\nonumber
\end{align}
%in the $r\to0$ and $r\to\infty$ limits, respectively.

For the reader's convenience, we plot in Fig.~\ref{Fig:Dymnikovaf} the function $f_{\scriptscriptstyle \text{D--AdS}}$ showing how it changes with $\ell$. As usual, depending on the value of this parameter two, one or no horizons can characterize the associated spacetime. In Fig.~\ref{Fig:DymnikovaAdSdS} we instead compare an extremal dSC-Dymnikova metric with an AdSC-Dymnikova one.
\begin{figure}
    \centering
    \includegraphics[width=\linewidth]{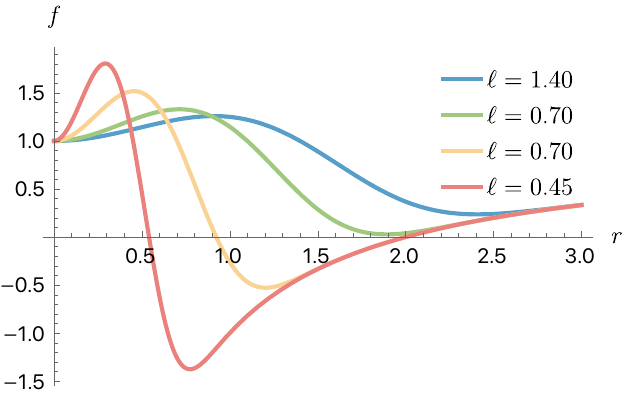}
    \caption{Metric function $f_{\scriptscriptstyle \text{D--AdS}}$ for different values of the regularization parameter $\ell$. Depending on its value, the metric can exhibit two, one, or no horizons. At large radii this metric quickly approaches the Schwarzschild solution.}
    \label{Fig:Dymnikovaf}
\end{figure}
\begin{figure}
    \centering
    \includegraphics[width=\linewidth]{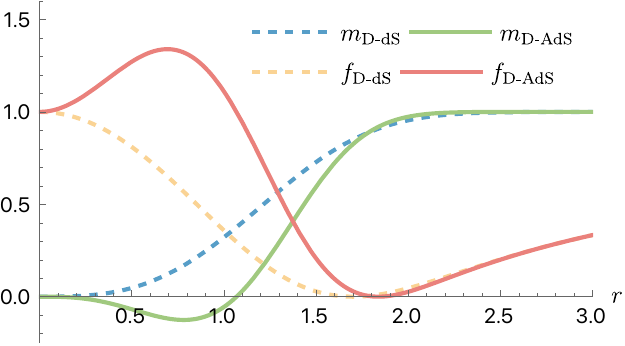}
    \caption{Comparison plot of the $f$ and $m$ functions for the dSC and AdSC-Dymnikova metrics. The regularization parameter has been fixed to $\ell_{\scriptscriptstyle \text{D--dS}}\approx 1.37$ and $\ell_{\scriptscriptstyle \text{D--AdS}}\approx 1.07$ so that both solutions exhibit a single, extremal horizon. Both metrics approach the Schwarzschild metric exponentially quickly at large radii, while they show large departures from it in the interior region.}
    \label{Fig:DymnikovaAdSdS}
\end{figure}

\subsubsection{Stress-energy tensor}
The structure of the SET is similar to the one in AdSC-Bardeen but with faster decay at large radii. The region with positive energy density is therefore more peaked as it has to compensate --- within a narrower region --- for the negative mass generated at the core. The density and pressure components are
\begin{align}
    \rho=
    &
    -p_{r}=
    \frac{Me^{-\frac{r^3}{\ell^3}\left(\frac{r^2-\ell^2}{r^2+\ell^2}\right)}}{4\pi \ell^3\left(r^2+\ell^2\right)^2}\times\left(3r^4+4r^2\ell^2-3\ell^4\right)\,,\nonumber\\
    p_{\theta}=
    &
    \frac{M e^{-\frac{r^3}{\ell^3}\left(\frac{r^2-\ell^2}{r^2+\ell^2}\right)}}{8\pi \ell^6\left(r^2+\ell^2\right)^4}\times\nonumber\\
    &
    \left(9r^{11}+24 r^9 \ell^2 -6 r^8\ell ^3 -2 r^7\ell^4 -24 r^6\ell^5 \right.\nonumber\\
    &
    \left.-24 r^5\ell^6-40 r^4 \ell^7+9 r^3 \ell^8 - 16 r^2 \ell ^ 9+ 6 \ell ^ {11}\right)\,.
\end{align}

It is easy to see that the energy density changes sign at the radius
\begin{equation}
r_{\rho}=\sqrt{\frac{\sqrt{13}-2}{3}}\ell,
\end{equation}
whereas the conditions $\rho+p_{\theta}\geq0$ and $p_{\theta}\geq0$ are violated for $0<r<1.15\ell$ and $0.49\ell<r<1.27\ell$.
In order to visualize all of the above, we plot the components of the SET for some exemplary value of $\ell$ in Fig.~\ref{Fig:DymnikovaSET}.
\begin{figure}
    \centering
    \includegraphics[width=\linewidth]{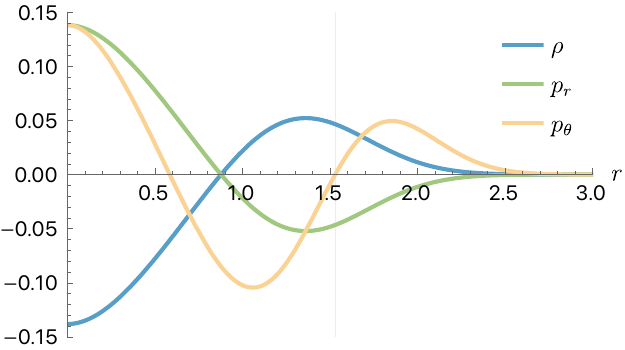}
    \caption{Stress-energy tensor components of the AdSC-Dymnikova solution for $\ell_{\scriptscriptstyle \text{D--AdS}}=1.20$, a solution of the horizonless kind. The SET components vanish exponentially quickly at sufficiently large radii. The strong energy condition is violated in the entire region to the left of the vertical line.}
    \label{Fig:DymnikovaSET}
\end{figure}

\subsection{More general Anti-de Sitter Core metrics}
\label{subsec:general}
We can generalize other metrics to obtain their AdSC version by further generalizing the function~\eqref{Eq:ChiRBH}
\begin{equation}\label{Eq:ChiRBH2}
    \chi\left(r,n,m,\epsilon,\kappa\right)=\frac{r^{n}-\epsilon\kappa^{n-m} \ell^{m}}{r^{n}+\epsilon\kappa^{n-m} \ell^{m}},\quad n\geq 1,\, \quad m \geq 1 ,
\end{equation}
where $\kappa$ is a positive constant with dimensions of length.

A generalization of the Fan-Wang metric is obtained by taking $\{n,m,\epsilon\}=\{1,1,1\}$
\begin{equation}\label{Eq:FanWang}
    f_{\scriptscriptstyle \text{FW--AdS}}=
    1-\frac{2M}{r}\left[\frac{r}{r+\ell}\right]^{3}\left(\frac{r-\ell}{r+\ell}\right)\, ,
\end{equation}
whereas for the Hayward black hole we could, for example, choose $\{n,m,\epsilon,\kappa\}=\{3,2,1,2M\}$, as in
\begin{equation}\label{Eq:Hayward}
    f_{\scriptscriptstyle \text{H--AdS}}=
    1-\frac{2M}{r}\left(\frac{r^{3}}{r^{3}+2M\ell^{2}}\right)\left(\frac{r^{3}-2M\ell^{2}}{r^{3}+2M\ell^{2}}\right) \, .
\end{equation}
%\jar{[[Question: Do you think it's useful to add more generalizations? What should we do with this section?]]}\ste{no}

Alternatively, it is certainly possible to produce metrics for which $f(r)\neq h(r)^{-1}$, allowing to disentangle the behaviour of the mass function from that of the redshift. This would allow to model objects with richer internal structures, such as those inspired by semiclassical physics~\cite{Arrechea:2023oax,Chen:2024ibc}, which fall within this category of metrics since their complicated stress-energy tensors do not satisfy $\rho+p_{r}=0$.

One such example can be easily proposed starting from the AdSC-Dymnikova metric and introducing the redshift factor, so to obtain a ``generalized Dymnikova" metric (GenD) with components
\begin{align}
    f_{\scriptscriptstyle \text{GenD--AdS}}=
    &
    \mathcal{F}(r)\left(1-\frac{2M}{r}\left[1-\exp{-\frac{r^{3}}{\ell^{3}}\left(\frac{r^{2}-\ell^{2}}{r^{2}+\ell^{2}}\right)}\right]\right)\, , \nonumber\\
    h_{\scriptscriptstyle \text{GenD--AdS}}=
    &
   \left(1-\frac{2M}{r}\left[1-\exp{-\frac{r^{3}}{\ell^{3}}\left(\frac{r^{2}-\ell^{2}}{r^{2}+\ell^{2}}\right)}\right]\right)^{-1}\, ,
\end{align}
where 
\begin{equation}
    \mathcal{F}(r)=1-\Phi_{0}e^{-r^3/\ell^3},
\end{equation}
and $\Phi_{0}$ is a positive constant taking values smaller than unity. 

The radial dependence of $\mathcal{F}$ prevents it from being fully absorbed in a redefinition of the $t$-coordinate, while its absence in the $rr$ component of the metric is what disentangle the behaviour of the mass function from that of the redshift. Near $r=0$ the $f$ function obeys
\begin{equation}
    f_{\scriptscriptstyle \text{GenD--AdS}}=1-\Phi_{0}+\frac{2M r^2\left(1-\Phi_{0}\right)}{\ell^3}+\Phi_{0}\frac{r^3}{\ell^3}+\order{r^4}\,,
\end{equation}
so that for $\Phi_{0}=0$ one would end up recovering the behaviour of the AdSC--Dymnikova solution (which is the same of the AdSC--Bardeen one~\eqref{Eq:Bardeenr0}). 

For the horizonless sub-family of metrics with $\ell>\ell_{\text{extr}}$, $\Phi_{0}<0$ generates smaller internal redshifts, while in the $\Phi_{0}\to1$ limit, it would instead take an arbitrarily long time for light rays to cross the object, effectively ``freezing" any signal and thus acting as a very efficient black hole mimicker. In this sense, this metric would reproduce some of the salient phenomenological features identified for some semiclassical star models in~\cite{Arrechea:2024nlp}.

\section{Phenomenological signatures}
\label{Sec:Pheno}

Perturbations provide a direct way to test how a given spacetime reacts to external disturbances. By studying the propagation of test fields on RBH and black hole mimicker backgrounds, we obtain quasinormal modes and echoes that serve as useful proxies for the qualitative features expected in realistic ringdown signals. This makes test perturbation analysis the natural first step in assessing whether models with different regular cores can be distinguished observationally.

In the test-field approach, one studies a passive probe --— here a field that interacts only through the spacetime metric (i.e.  with no additional couplings to curvature or matter) --— so that the background geometry and any matter sources remain fixed. After separating the angular dependence in spherical harmonics, the field obeys a Schrödinger-like wave equation
\begin{equation}\label{eq: master}
    \frac{\partial^2 \psi}{\partial t^2}-\frac{\partial^2 \psi}{\partial r_{*}^2}+V_{s}(r)\psi=0\,,
\end{equation}
whose effective potential $V_{s}(r)$ is completely determined by the background functions $f(r)$ and $h(r)$ in~\eqref{Eq:SphSymmetric},
and the tortoise coordinate is defined via $dr_{*}/dr=\sqrt{f/h}$. In particular, for $f(r)={h(r)}^{-1}=1-2m(r)/r$ one has
\begin{equation}\label{eq: effpot}
    V_{s}=f(r) \left[\frac{l(l+1)}{r^2}+\frac{2(1-s^2)m(r)}{r^3}-(1-s)\frac{2m'(r)}{r^2} \right]\,,
\end{equation}
where $l$ is the harmonic index of the perturbation. This expression generalizes the Schwarzschild Regge-Wheeler potential~\cite{FrolovNovikov1998} for arbitrary functions $f(r)$ and $m(r)$. Equation~\eqref{eq: master} can be solved directly in the frequency domain, assuming a harmonic $t$-dependence in the field and finding the modes obeying specific boundary conditions, or in the time domain, where the field initially describes a small perturbation whose evolution in $t$ and $r_{*}$ is calculated as it probes the background geometry. Because the field is a test perturbation, it decouples from the gravitational sector. Furthermore, the master equation for perturbations of different spin $s$ differs only through the spin-dependent terms in $V_{s}$~\eqref{eq: effpot}.

%Because the field is a test perturbation, it decouples from the gravitational sector; this guarantees that, even for exotic interiors, the exterior behavior is governed solely by the outer light ring and the imposed boundary conditions. Working in spherical symmetry further simplifies matters: the master equation for perturbations of different spin $s$ differs only through the spin-dependent terms in $V_{s}$~\eqref{eq: effpot}.

When working in the frequency domain, the real part of the fundamental (or lowest frequency) QNM sets the characteristic ringing frequency, while the imaginary part fixes the damping time. In the frequency domain, we will consider metrics with and without horizons, taking in each case boundary conditions (defined below) adapted to the situation at hand.
In the time domain, however, we will focus on horizonless configurations, where the corresponding solution will exhibit late-time ``echoes" due to the presence of a second peak in the potential. The echo delay is essentially the light-crossing time of the internal cavity between the two peaks of the potential, i.e.,
\begin{equation}\label{eq: crosstime}
\tau_{\text{ph}}=2\int_{0}^{r_{\text{ph}}}\sqrt{\frac{h}{f}}dr\, ,
\end{equation}
and the Fourier transform of the corresponding signal reproduces the low-lying QNMs, providing a direct bridge between the two analyses. Throughout this work we therefore solve the test-field equation~\eqref{eq: master} for each of the dSC and AdSC models presented in Section~\ref{Sec:RBHs}.

\subsection{Quasinormal modes}
The complex fundamental frequencies characterizing the solutions of \eqref{eq: master} are obtained with a direct-integration (shooting) method as developed by Chandrasekhar and Detweiler \cite{Chandrasekhar:1975zza,Chandrasekhar:1991fi}. Using Frobenius expansions to derive the corresponding boundary conditions, we integrate the radial Schrödinger-like equation outward from the inner boundary and inward from a large radius, adjusting the trial frequency until the two solutions and their first radial derivatives match smoothly at an intermediate point, which implies a vanishing Wronskian. For black holes, the inner expansion is taken at the event horizon with purely ingoing boundary conditions. For horizonless configurations, the integration begins instead at the regular center, where a Dirichlet condition is imposed. In both cases, the accuracy of our results is verified by checking that the fundamental QNM frequency does not change upon varying the matching radius.

Near the event horizon, the field amplitude
is expanded as a power series in $(r-r_{h})$
\begin{equation}\label{eq: nearhor}
    \Psi(r\sim r_{h})\sim e^{-i\omega r_{*}}\sum_{i=0}^{\infty}p_{i}(r-r_{h})^{i}\, ,
\end{equation}
where $p_{i} $ are constant coefficients, and the infinite sum is truncated so that the boundary conditions converge to the desired accuracy. As a result, the wave equation itself is expanded order by order. This yields a recursion relation that expresses higher-order coefficients in terms of the leading one, thereby generating the Frobenius solution consistent with purely ingoing boundary conditions. 

A similar expansion is constructed at large radii, where the solution is expressed in powers of ${1}/{r}$ and enforces purely outgoing behavior, i.e.,
\begin{equation}\label{eq: nearinf}
    \Psi(r\to\infty)\sim e^{i\omega r_{*}}\sum_{i=0}^{\infty}\frac{q_{i}}{r^{i}}\, .
\end{equation}
Numerical integration is then performed: one branch evolves outward from the horizon to a chosen matching point, and the other inward from infinity to the same location. The complex frequency is adjusted until the Wronskian of the two solutions vanishes, guaranteeing smooth matching.
For horizonless spacetimes the procedure is identical, except that the inner Frobenius expansion is taken at the regular center. There the boundary condition is \cite{Macedo:2016wgh} 
\begin{equation}\label{eq: nearcenter}
    \Psi(r\sim 0)\sim r^{l} \sum_{i=0}^{N'_{0}} \tilde{p}_{i}\ r^{i}\, ,
\end{equation}
where $\tilde{p}_{i}$ are constant coefficients, and the upper limit of the summation is again chosen such that the boundary conditions converge to the desired accuracy. Integration proceeds outward from the center, and the matching condition with the inward solution from infinity is imposed exactly as in the black-hole case.

%, which ensures the robustness of the spectrum and follows the standard practice established in previous studies of RBHs. 
In what follows we present the results of the method for the Bardeen and Dymnikova families of metrics considered previously.
\subsubsection{Bardeen}
For the Bardeen family of solutions, we show in Fig.~\ref{fig:BardeenQNMs} a comparison 
between a dSC and an AdSC Bardeen metric, with and without horizons. In the black hole case, the imaginary parts of the frequency slightly decrease with $\ell$.
It is evident that the largest deviations are present close to extremality (see Fig.~\ref{Fig:BardeenAdSdS} for a comparison of the metrics of two near-extremal solutions), where the differences in the cores are maximized and percolate outside the horizons. For the black hole mimicker case, the real part shows growing discrepancies for large $\ell$ values, while the imaginary part always increases, indicating a less effective trapping of light rays as the object is made less compact. 
\begin{widetext}
  \begin{center}
    \includegraphics[width=0.90\textwidth,keepaspectratio]{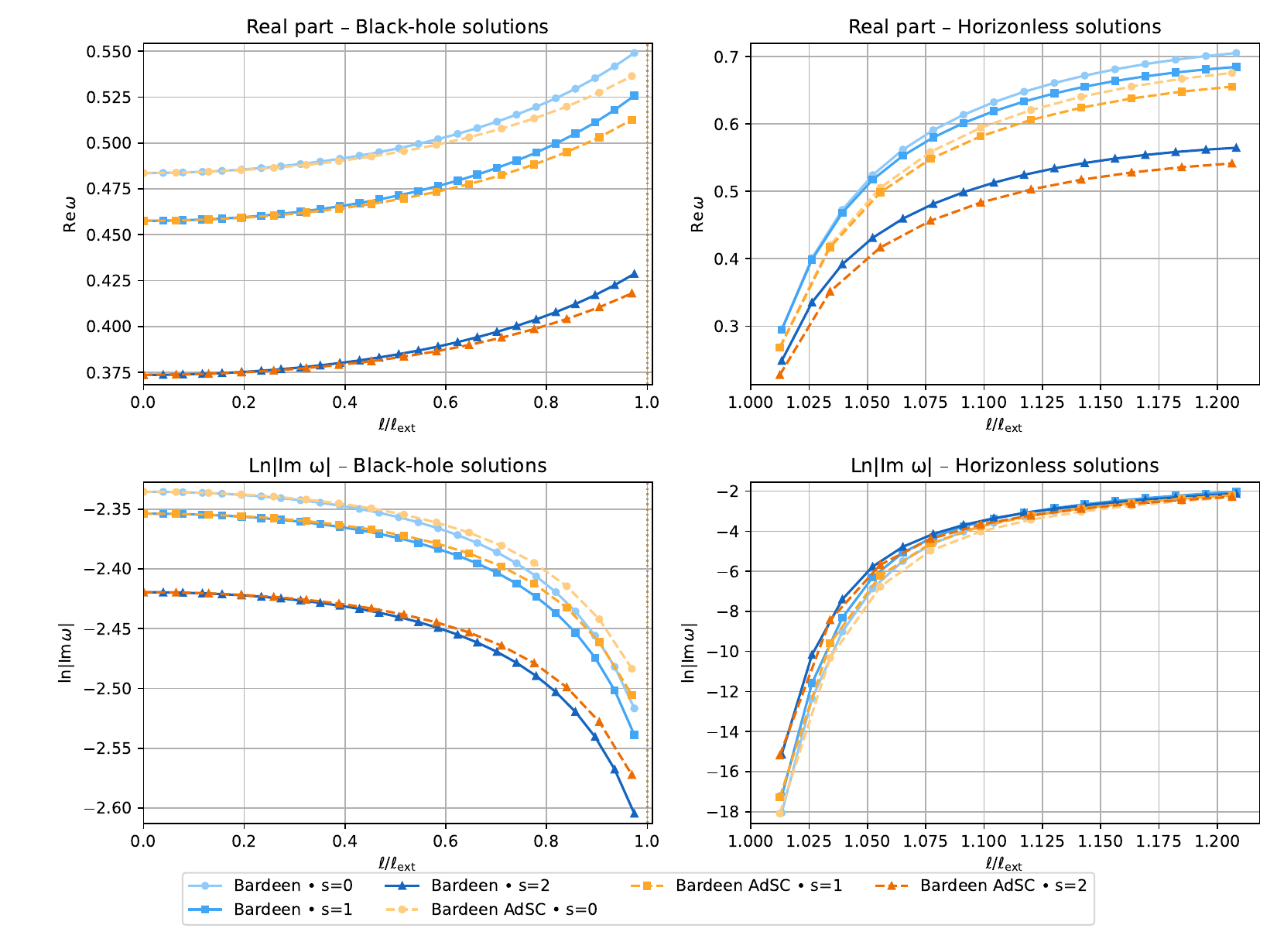}
    \captionof{figure}{Fundamental quasinormal mode frequencies for dSC-Bardeen and AdS-Bardeen solutions. The horizontal axis shows the regularization parameter normalized by its extremal value, $\ell/\ell_{\text{ext}}$. Top panels: real part of the frequencies, for black-hole solutions (left) and horizonless solutions (right). Bottom panels: logarithm of the absolute value of the imaginary part of the frequencies, for black-hole solutions (left) and horizonless solutions (right). The logarithmic scale is adopted because the damping rates (imaginary parts) are very small and close to each other, making their differences visible only after rescaling.  Shades of blue correspond to dSC-Bardeen, shades of orange to AdSC-Bardeen, with different shades denoting spin $s=\ 0,\ 1,\ 2$.}
    \label{fig:BardeenQNMs}
  \end{center}
\end{widetext}

\subsubsection{Dymnikova}
For the Dymnikova family we apply the same direct-integration scheme outlined above, modulo one technical difference that arises in the outer expansion~\eqref{eq: nearinf}. The Dymnikova spacetime leads to computationally heavy coefficients in the large-$r$ expansions. Since the metric approaches the Schwarzschild metric exponentially quickly, we can safely neglect exponentially-decaying terms in Eq.~\eqref{eq: master} that encode deviations with respect to the Regge-Wheeler equation in Schwarzschild. This simplification preserves the correct asymptotic structure of the modes while greatly reducing the computational cost of generating the coefficients. Apart from this modification, the procedure mirrors the Bardeen case for both black hole and horizonless configurations. 

In Fig.~\ref{fig:DymnikovaQNMs} we show a comparison between a dSC and an AdSC-Dymnikova metric, with and without horizons. As in the Bardeen case, the largest discrepancies arise in the black-hole branch close to extremality (see Fig.~\ref{Fig:DymnikovaAdSdS} for nearly-extremal solutions). In this regime, both the real and the imaginary parts display markedly different trends for the de Sitter and Anti–de Sitter cores. By comparison, the horizonless branches exhibit growing discrepancies in the real part for large $\ell$ values, similar to the Bardeen case. 
\begin{widetext}
  \begin{center}
    \includegraphics[width=0.90\textwidth,keepaspectratio]{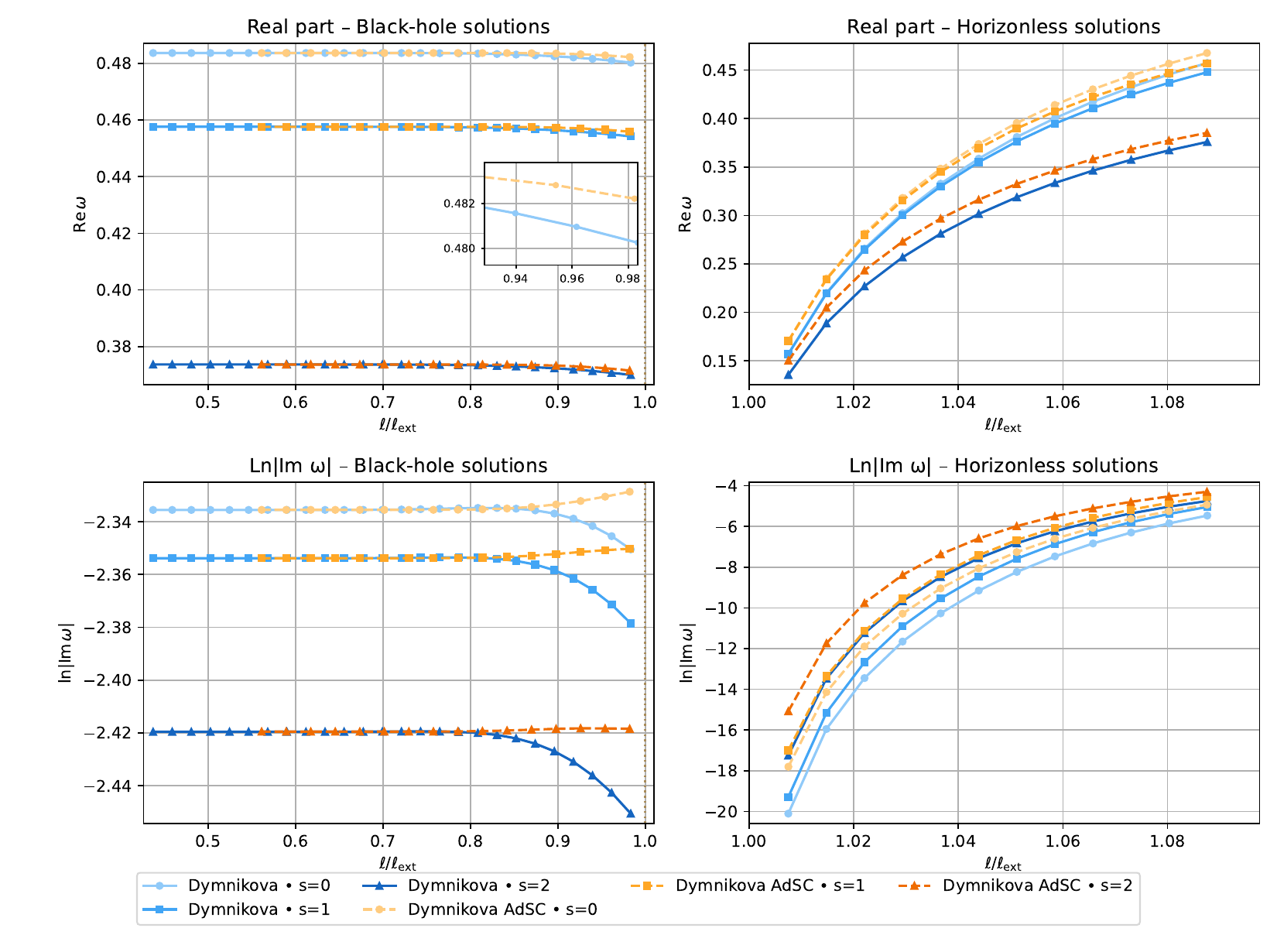}
    \captionof{figure}{Fundamental quasinormal mode frequencies for dSC-Dymnikova and AdSC-Dymnikova solutions. The horizontal axis shows the regularization parameter normalized by its extremal value, $\ell/\ell_{\mathrm{ext}}$. Top panels: real part of the frequencies, for black-hole solutions (left) and horizonless solutions (right). Bottom panels: logarithm of the absolute value of the imaginary part of the frequencies, for black-hole solutions (left) and horizonless solutions (right). The logarithmic scale is adopted because the damping rates (imaginary parts) are very small and close to each other, making their differences visible only after rescaling. Shades of blue correspond to dSC-Dymnikova, shades of orange to AdSC-Dymnikova, with different shades denoting spin $s=0,1,2$. A zoomed inset is included in the top-left panel to highlight the near-extremal region of the spin-$0$ branch, where the real parts of the two cores exhibit clear deviations.}  % needs \usepackage{caption}
    \label{fig:DymnikovaQNMs}
  \end{center}
\end{widetext}

\subsection{Time-domain analysis}
In the time-domain analysis, we deliberately remove the horizon by pushing the regularization parameter above its extremal value, selecting two representative mass families --- Bardeen-type and Dymnikova-type --- and, for each of them, tuning \(\ell\) so that the local maximum of the effective potential coincides. Since this maximum corresponds to the outer photon sphere (the unstable light ring), aligning it across models ensures that differences in the time separation between echoes are due to the different trapping of light rays by the inner core rather than to trivial shifts of the exterior barrier.
The master equation~\eqref{eq: master} is then evolved in the time domain. We initialize the system with a Gaussian wave packet at rest
\begin{equation}\label{eq: initialpacket}
    \Psi(t=0,r_{*})= A \exp(-\frac{(r_{*}-r_{*0})^2}{\sigma^2})\,, \quad \dot{\Psi}(t=0,r_{*})=0\,,
\end{equation}
where the dot denotes the derivative with respect to the $t$ coordinate.
The subsequent signal is entirely generated by the scattering of this pulse on the effective potential. Numerical evolution is performed with a fourth–order Runge–Kutta integrator on a uniform $(t,r_{*})$ grid, subject to Courant stability. At the inner boundary ($r=0$ for horizonless models) we impose a total reflection by enforcing a Dirichlet condition, $\Psi=0$, which in spherical symmetry is necessary to ensure regularity and is standard in echo studies of horizonless compact objects. At large radius we impose a Sommerfeld condition $(\partial_{t}+\partial_{r_{*}})\Psi=0$, ensuring purely outgoing absorption at infinity. 
In what follows we present the results of the method for the two families so far considered.

\subsubsection{Bardeen}

For Bardeen–type black hole mimickers, we present the evolution of the spin-$2$ perturbation in Fig.~\ref{fig:waveformB}. We clearly see a sizeable shift in the echo response when examining the waveform at a fixed physical radius. The signals measured by the observer are visibly distinct in the dS–Core and AdSC cases. In particular, there is a clear time delay between the peaks of the reflected waves: the AdSC peaks take less time to probe the interior and reach the observer compared to the dSC peaks.
\begin{widetext}
  \begin{center}
    \includegraphics[width=0.90\textwidth,keepaspectratio]{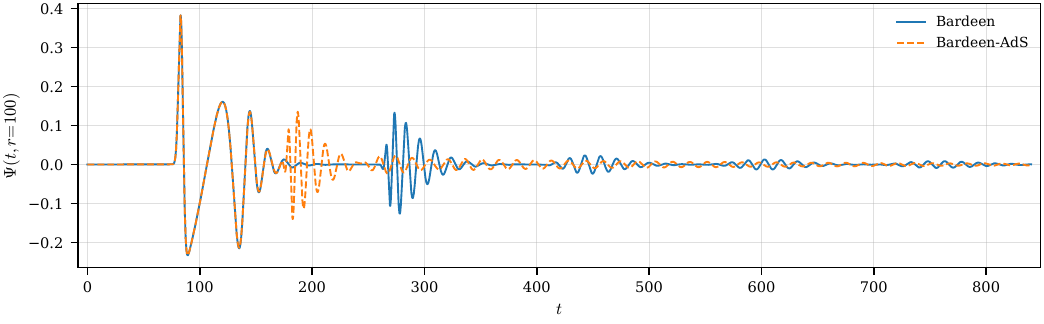}
    \captionof{figure}{Time–domain evolution of a spin-2 test field on horizonless dSC-Bardeen (blue) and AdSC-Bardeen (orange) backgrounds. The parameters \(\ell\) are chosen so that the first local maximum of the Regge–Wheeler potential occurs at the same radius \(r_{\max}\) in both cases ($\ell_{\scriptscriptstyle \text{B}}=0.7728\ $, $\ell_{\scriptscriptstyle \text{B-AdS}}=0.4810$). Shown is the waveform \(\Psi(t, r=100)\) as measured by an observer at a fixed physical radius \(r=100\), chosen well outside the compact object. The evolution starts from a stationary Gaussian packet centered outside \(r_{\max}\), with Dirichlet regularity imposed at the origin and a Sommerfeld absorbing condition at large \(r_*\). The echo trains clearly differ between the de-Sitter-core (Bardeen) and AdSC (Bardeen–AdS) models, with the time delay between successive reflected pulses depending on the interior structure.}  % needs \usepackage{caption}
    \label{fig:waveformB}
  \end{center}
\end{widetext}
One can explicitly calculate this time delay between the two cases using \eqref{eq: crosstime}. Alternatively, for a more qualitative explanation of this difference, one can look at the shape of the effective potentials associated to the two solutions Fig.~\ref{fig:potB}.

\subsubsection{Dymnikova}

For the Dymnikova class of horizonless spacetimes we obtain a qualitatively similar behavior. The waveform at a fixed physical radius $r=100$ shows differences between the dS–Core and AdS–Core cases. The contrast between the two is smaller than in the Bardeen model, but it is nevertheless visible in the time delay between successive reflected peaks, with the AdS–Core version producing slightly earlier echoes. 
\begin{widetext}
  \begin{center}
    \includegraphics[width=0.90\textwidth,keepaspectratio]{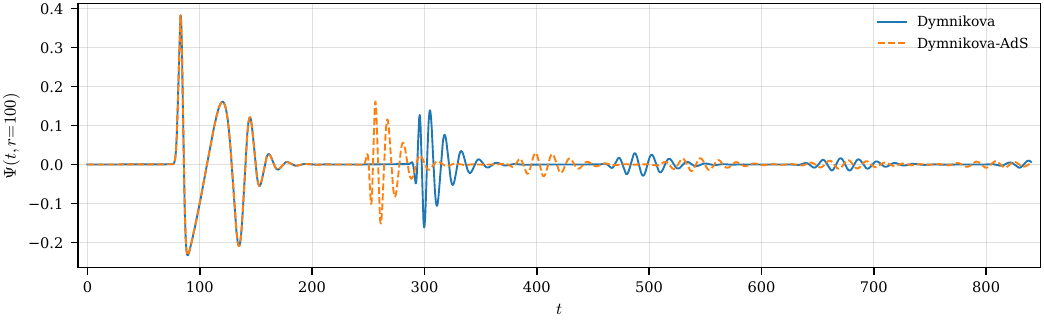}
    \captionof{figure}{Time–domain evolution of a spin-2 test field on horizonless dSC-Dymnikova (blue) and AdSC-Dymnikova (orange) backgrounds. The regularization parameters are tuned so that the outer potential peak lies at the same radius \(r_{\max}\) for both spacetimes ($\ell_{\scriptscriptstyle \text{D}}=1.3765\ $, $\ell_{\scriptscriptstyle \text{D-AdS}}=1.0713$). Shown is the waveform \(\Psi(t, r=100)\) as measured by an observer at a fixed physical radius \(r=100\), chosen well outside the compact object. The evolution starts from a stationary Gaussian packet centered outside \(r_{\max}\), with Dirichlet regularity imposed at the origin and a Sommerfeld absorbing condition at large \(r_*\). As in the Bardeen case, the echoes differ between the dSC-Dymnikova and AdSC-Dymnikova models, although here the contrast is smaller, with the AdSC solution producing slightly earlier pulses.}  % needs \usepackage{caption}
    \label{fig:waveformD}
  \end{center}
\end{widetext}

Similarly, this difference can be traced back to the shape of the effective potentials Fig.~\ref{fig:potD}. 
\begin{figure}
    \centering
    \includegraphics[width=\linewidth]{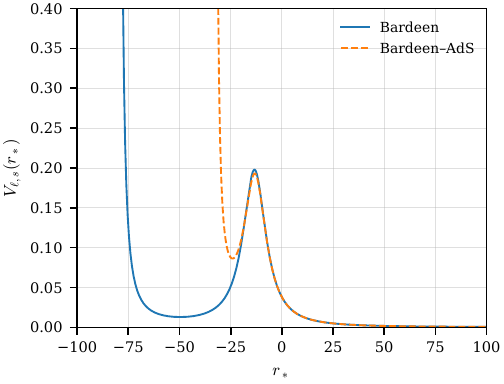}
    \caption{Regge--Wheeler potentials \(V_{\ell,s}(r_\ast)\) as functions of the tortoise coordinate \(r_\ast\) for the horizonless dSC-Bardeen and AdSC-Bardeen backgrounds. Parameters: \(M=1\), \(l=2\), \(s=2\), \(\ell_{\scriptscriptstyle \text{B-dS}}=0.7728\), \(\ell_{\scriptscriptstyle \text{B-AdS}}=0.4810\).}
    \label{fig:potB}
\end{figure}
\begin{figure}
    \centering
    \includegraphics[width=\linewidth]{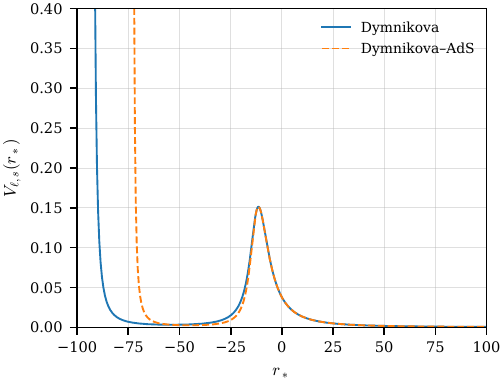}
    \caption{Regge--Wheeler potentials \(V_{\ell,s}(r_\ast)\) as functions of the tortoise coordinate \(r_\ast\) for the horizonless dSC-Dymnikova and AdSC-Dymnikova backgrounds. Parameters: \(M=1\), \(l=2\), \(s=2\), \(\ell_{\scriptscriptstyle \text{D-dS}}=1.3765\), \(\ell_{\scriptscriptstyle \text{D-AdS}}=1.0713\).}
    \label{fig:potD}
\end{figure}

The analysis is indicating evident differences in the echoes observed for each family between the $\mbox{dSC}$ and $\mbox{AdSC}$ cases. Clearly, this difference stems from the behavior of effective potentials deep inside the compact object, even though the potentials are identical in the $r\to \infty$ limit. The difference in the distances between the reflective barrier at $r=0$ and the outer photon sphere can be traced back to the difference between the $f(r)$ and $h(r)$ functions in both models, which the reader can revisit in Figures~\ref{Fig:BardeenAdSdS} and~\ref{Fig:DymnikovaAdSdS}. The presence of the AdS core makes the $f(r)$ function take larger values than its dS counterpart near the cores of the Bardeen and Dymnikova metrics. Similarly, the negative mass core makes $h(r)$  reach, on average, smaller absolute values in the AdS case. Consequently, we can expect the crossing time~\eqref{eq: crosstime} to be smaller for the AdS case, and this feature is reflected clearly in the effective potentials from Figures~\ref{fig:potB} and~\ref{fig:potD}. 
Finally, notice that this is a consequence of the prescription that we have devised to turn de Sitter cores into Anti-de Sitter ones and, as such, is a model-dependent feature. In any case, this analysis highlights some observational differences that could be used to rule out or distinguish models in future observations. 

\section{Conclusions}
\label{Sec:Conclusions}

In this work, we have explored the possibility that regular black holes and black hole mimickers may host Anti-de Sitter cores rather than the more commonly assumed de Sitter interiors. We exposed the connection between the presence of regions of spacetime with large energies/pressures and the emergence of an effectively vacuum equation of state with negative or positive cosmological constant, respectively. Following this thread, and motivated by recent semiclassical considerations showing that vacuum polarization in ultra-compact configurations can lead to effective negative energy densities, we proposed a general prescription to obtain AdSC counterparts of known regular black hole metrics and applied it to the Bardeen and Dymnikova families. A generic feature of these models is the presence of a negative mass core compensated by a surrounding positive-mass layer, closely paralleling semiclassical stellar models. While we incorporated these features via the introduction Anti-de Sitter cores, the models presented here should be understood as particular examples of a broader class of potential objects sustained by the same underlying physical principles. 

After introducing a general prescription to obtain these metrics, we investigated their phenomenological signatures via test-field perturbations. Our analysis of quasinormal modes revealed systematic deviations between AdSC and dSC geometries, particularly in the near-extremal regime where observational imprints could be most pronounced. In the horizonless case, we found that the AdS interiors modify the structure of echoes, altering both their timing and waveform. These results highlight the potential of gravitational-wave observations --- especially precise measurements of ringdowns and echoes --- to probe not only the presence of horizons but also the inner structure of ultra-compact objects. 

It is worth noting that for the AdSC solutions considered here, short-delay echoes are found. This is at odds with the phenomenology reported in~\cite{Arrechea:2024nlp} for semiclassical stars (also with negative energy density cores): for those objects, the delays were found to be so long as to be effectively undetectable. This is a purely relativistic effect due to the huge interior crossing times required for light rays to traverse their structures. The time delay between echoes and the efficiency with which they become trapped is controlled by the depth of the effective potential $V_s(r)$ appearing in \eqref{eq: master}.

%This outcome highlights that echo delays, contrary to common wisdom, are not simply related to a naive light-crossing time, nor to the specific geometry of the objects' cores. Rather, they are controlled by the depth of the effective potential $V_s(r)$ in see \eqref{eq: master} at their center.

The horizonless AdSC solutions we generated from the most common dSC RBH metrics share with them a ``shallow'' effective potential. This is not surprising: in these one-parameter families of metrics, the potential depth is inversely proportional to the regularization parameter $\ell$, and the latter must be macroscopic for the geometry to be horizonless. Mimicking geometries with ``deep'' cores, such as the semiclassical stars studied in~\cite{Arrechea:2024nlp}, requires breaking this direct relation, e.g., by considering families of metrics characterized by more than one parameter, or by introducing additional redshift-controlling functions, as briefly discussed in~\ref{subsec:general}.

%From a broader perspective, our findings suggest that the assumption of dS interiors in regular black holes is not unique, but rather one branch of a more general paradigm where both signs of the effective vacuum energy are possible. Which scenario is realized in nature may depend sensitively on the balance between density and pressure at Planckian scales. In this sense, AdSC mimickers provide a fertile framework to bridge semiclassical gravity, modified gravity models, and astrophysical observations.  

Several avenues remain open. On the theoretical side, it would be important to connect the effective stress-energy tensors found here with explicit semiclassical or quantum-gravity-inspired models, to assess the stability of AdSC solutions, and to clarify their possible dynamical formation channels. On the observational side, it will be crucial to quantify to what extent future gravitational-wave detectors, with improved sensitivity in the ringdown regime, could distinguish between dS- and AdSC signatures.  

Ultimately, the exploration of AdSC regular black holes invites us to revisit the nature of compact objects with fresh eyes. If regular black holes and black hole mimickers exist, their internal structure may be far richer than previously thought, and gravitational waves may provide the key to unveiling it. As with a black box that we shake and listen to in order to infer its contents, the ringdown may bring the cores of regular black holes, and of black hole mimickers, within observational reach. 

%Only time will tell.

%============================================================
\begin{acknowledgments}
	The authors thank Johanna Borissova and Matt Visser for insightful discussions.
\end{acknowledgments}
%============================================================
\bibliographystyle{apsrev4-1}
	\bibliography{main}

%============================================================
\end{document}